\documentclass{JHEP3}
\usepackage{amssymb}

\def \be {\begin{equation}}
\def \ee {\end{equation}}

\def \bea {\begin{eqnarray}}
\def \eea {\end{eqnarray}}
\def \nn {\nonumber}

\def \a {\alpha}
\def \b {\beta}
\def \g {\gamma}
\def \G {\Gamma}
\def \d {\delta}

\def \m {\mu}
\def \n {\nu}
\def \k {\kappa}

\def \s {\sigma}
\def \r {\rho}
\def \o {\omega}
\def \O {\Omega}
\def \th {\theta}
\def \Th {\Theta}

\def \t {\tau}
\def \dag {\dagger}
\def \p {\partial}

\def\bd{\begin{document}}
\def\ed{\end{document}}
\def\nn{\nonumber}
\def\bea{\begin{eqnarray}}
\def\eea{\end{eqnarray}}
\let\bm=\bibitem
\let\la=\label

\def\N{{\cal N}}
\def\sst{\scriptscriptstyle}
\def\thetabar{\bar\theta}
\def\Tr{{\rm Tr}}
\def\one{\mbox{1 \kern-.59em {\rm l}}}

\def\a{\alpha}      \def\da{{\dot\alpha}}
\def\b{\beta}       \def\db{{\dot\beta}}
\def\c{\gamma}  \def\C{\Gamma}  \def\cdt{\dot\gamma}
\def\d{\delta}  \def\D{\Delta}  \def\ddt{\dot\delta}
\def\e{\epsilon}        \def\vare{\varepsilon}
\def\f{\phi}    \def\F{\Phi}    \def\vvf{\f}
\def\h{\eta}
\def\k{\kappa}
\def\l{\lambda} \def\L{\Lambda}
\def\m{\mu} \def\n{\nu}
\def\o{\omega}
\def\P{\Pi}
\def\r{\rho}
\def\s{\sigma}  \def\S{\Sigma}
\def\t{\tau}
\def\th{\theta} \def\Th{\Theta} \def\vth{\vartheta}
\def\X{\Xeta}
\def\z{\zeta}
\def\w{\wedge}
\def\u{\underline}
\def\hs{\hspace}

\def\cA{{\cal A}} \def\cB{{\cal B}} \def\cC{{\cal C}}
\def\cD{{\cal D}} \def\cE{{\cal E}} \def\cF{{\cal F}}
\def\cG{{\cal G}} \def\cH{{\cal H}} \def\cI{{\cal I}}
\def\cJ{{\cal J}} \def\cK{{\cal K}} \def\cL{{\cal L}}
\def\cM{{\cal M}} \def\cN{{\cal N}} \def\cO{{\cal O}}
\def\cP{{\cal P}} \def\cQ{{\cal Q}} \def\cR{{\cal R}}
\def\cS{{\cal S}} \def\cT{{\cal T}} \def\cU{{\cal U}}
\def\cV{{\cal V}} \def\cW{{\cal W}} \def\cX{{\cal X}}
\def\cY{{\cal Y}} \def\cZ{{\cal Z}}
\def\bo {\bar{\o}}

\def\ua{\underline{\alpha}} \def\ubb{\underline{\beta}}
\def\ug{\underline{\gamma}}
\def\ub{\underline{\phantom{\alpha}}\!\!\!\beta}
\def\uc{\underline{\phantom{\alpha}}\!\!\!\gamma}
\def\um{\underline{\mu}} \def\un{\underline{\nu}}
\def\ud{\underline\delta}
\def\ue{\underline\epsilon}
\def\una{\underline a}\def\unA{\underline A}
\def\unb{\underline b}\def\unB{\underline B}
\def\unc{\underline c}\def\unC{\underline C}
\def\und{\underline d}\def\unD{\underline D}
\def\une{\underline e}\def\unE{\underline E}
\def\unf{\underline{\phantom{e}}\!\!\!\! f}\def\unF{\underline F}
\def\unm{\underline m}\def\unM{\underline M}
\def\unn{\underline n}\def\unN{\underline N}
\def\unp{\underline{\phantom{a}}\!\!\! p}\def\unP{\underline P}
\def\unq{\underline{\phantom{a}}\!\!\! q}
\def\unQ{\underline{\phantom{A}}\!\!\!\! Q}
\def\unH{\underline{H}}
\def\ul{\underline}

\def\As {{A \hspace{-6.4pt} \slash}\;}
\def\bs {{b \hspace{-6.4pt} \slash}\;}
\def\Ds {{D \hspace{-6.4pt} \slash}\;}
\def\ds {{\del \hspace{-6.4pt} \slash}\;}
\def\ss {{\s \hspace{-6.4pt} \slash}\;}
\def\ks {{ k \hspace{-6.4pt} \slash}\;}
\def\ps {{p \hspace{-6.4pt} \slash}\;}
\def\pas {{{p_1} \hspace{-6.4pt} \slash}\;}
\def\pbs {{{p_2} \hspace{-6.4pt} \slash}\;}

\def\Fh{\hat{F}}
\def\Vh{\hat{V}}
\def\Xh{\hat{X}}
\def\ah{\hat{a}}
\def\xh{\hat{x}}
\def\yh{\hat{y}}
\def\ph{\hat{p}}
\def\xih{\hat{\xi}}

\def\psit{\tilde{\psi}}
\def\Psit{\tilde{\Psi}}
\def\tht{\tilde{\th}}

\def\At{\tilde{A}}
\def\Qt{\tilde{Q}}
\def\Rt{\tilde{R}}
\def\Nt{\tilde{N}}

\def\at{\tilde{a}}
\def\st{\tilde{s}}
\def\ft{\tilde{f}}
\def\pt{\tilde{p}}
\def\qt{\tilde{q}}
\def\vt{\tilde{v}}
\def\nt{\tilde{n}}

\def\delb{\bar{\partial}}
\def\bz{\bar{z}}
\def\bD{\bar{D}}
\def\bB{\bar{B}}
\def\bo {\bar{\o}}

\def\bk{{\bf k}}
\def\bl{{\bf l}}
\def\bp{{\bf p}}
\def\bq{{\bf q}}
\def\br{{\bf r}}
\def\bx{{\bf x}}
\def\by{{\bf y}}
\def\bR{{\bf R}}
\def\bV{{\bf V}}
\def\bd{\begin{document}}

\def\ed{\end{document}}

\def\d{\delta}\def\D{\Delta}\def\ddt{\dot\delta}

\def\p{\partial} \def\del{\partial}
\def\xx{\times}
\def\uno{\mbox{1 \kern-.59em {\rm l}}}

\def\trp{^{\top}}
\def\inv{^{-1}}
\def\dag{{^{\dagger}}}
\def\pr{\prime}

\def\rar{\rightarrow}
\def\lar{\leftarrow}
\def\lrar{\leftrightarrow}

\def\cw{{\cal W}}
\def\cz{{\cal Z}}
\def\tcm{\tilde{\cal M}}
\def\sgn{{\rm sgn}}
\def\sd {d^{4|4}}
\def\lan{\langle}
\def\ran{\rangle}

\title{ Hidden Conformal Symmetry of Extremal Black Holes}
\author{Bin Chen\\
Department of Physics,\\
and State Key Laboratory of Nuclear Physics and Technology,\\
and Center for High Energy Physics,\\
Peking University,\\
Beijing 100871, P.R. China\\

\email{bchen01@pku.edu.cn}}

\author{Jiang Long, Jia-ju Zhang\\
Department of Physics,\\
and State Key Laboratory of Nuclear Physics and Technology,\\
Peking University,\\
Beijing 100871, P.R. China\\
\email{longjiang0301@gmail.com, jjzhang@pku.edu.cn}}

\date{\today}

\abstract{We study the hidden conformal symmetry of extremal
black holes. We introduce a new set of conformal coordinates to
write the $SL(2,R)$ generators. We find that the Laplacian of the
scalar field in many extremal black holes, including Kerr(-Newman),RN,
 warped AdS$_3$ and null warped black holes, could be written in terms
of the $SL(2,R)$ quadratic Casimir. This suggests that there exist
dual conformal field theory (CFT) descriptions of these black holes. From the conformal
coordinates, the temperatures of the dual CFTs could be read
directly. For the extremal black hole, the Hawking temperature is
vanishing. Correspondingly, only the left (right) temperature of the
dual CFT is nonvanishing and the excitations of the other sector
are suppressed. In the probe limit, we compute the scattering
amplitudes of the scalar off the extremal black holes and find
perfect agreement with the CFT prediction.
 }

\newpage
\begin{document}

\section{Introduction}\label{sec-intro}

The AdS/CFT correspondence\cite{AdSCFT} states that the quantum
gravity in anti-de-Sitter(AdS) spacetime is dual to a conformal
field theory(CFT)  at the AdS boundary. This discovery opens a new
window to study the quantum gravity. In this correspondence, the
black hole asymptotic to the AdS spacetime could be described by the
CFT at a finite temperature. In \cite{AndyWei}, it was conjectured
that even asymptotically flat four-dimensional Kerr black hole may have a
holographic two-dimensional CFT description. This conjecture was first set up for
extremal and near-extremal Kerr black hole by studying their
near-horizon geometry\cite{AndyWei,
matsuo,Castro:2009jf,{Bardeen:1999px}} and super-radiant
scattering\cite{Bredberg:2009pv,Hartman:2009nz,Cvetic:2009jn,ChenChu,{Becker:2010jj}}.
Very recently, this Kerr/CFT correspondence was generalized to the 
generic nonextremal Kerr black hole\cite{Castro:2010fd}. The key
ingredient is  the hidden conformal symmetry of the Kerr black hole.

For a generic nonextremal Kerr black hole, the hidden conformal
symmetry is realized by introducing a set of  conformal
coordinates:
  \bea
\o^+&=&\sqrt{\frac{r-r_+}{r-r_-}}e^{2\pi T_R\phi+2n_R t},\nn\\
\o^-&=& \sqrt{\frac{r-r_+}{r-r_-}}e^{2\pi T_L\phi+2n_L t},\label{confcoord0}\\
y&=&\sqrt{\frac{r_+-r_-}{r-r_-}}e^{\pi (T_L+T_R)\phi+(n_L+n_R)t},\nn
\eea where $r_\pm$ are the horizons of the black hole.  With
$(\o^\pm,y)$, one can locally define two sets of vector fields
satisfying the $SL(2,R)$ Lie algebra. More importantly, the scalar
Laplacian in the low-frequency limit could be written as the
$SL(2,R)$ quadratic Casimir, showing  hidden $SL(2,R)\times
SL(2,R)$ symmetries. Even though these $SL(2,R)$ symmetries are not
globally defined and are broken by the angular identification
$\phi\sim \phi +2\pi$, they act on the solution space, and determine the
form of the scattering amplitudes. It is also remarkable that the
conformal coordinates $\o^\pm$ encode the information of the
temperatures of the dual CFT, as a result of the Unruh effect. The
existence of this hidden conformal symmetry is essential to set up a
CFT dual to nonextremal Kerr black hole. The study of hidden
conformal symmetry has been generalized to other kinds of black hole
and applied to compute the real-time correlators in Kerr/CFT, see
\cite{Krishnan:2010pv}-\cite{Matsuo:2010in}.

However, the conformal coordinates (\ref{confcoord0}) do not make
sense for extremal black holes. In this case, as $r_+=r_-$ the
coordinate $y$ is simply zero and not well-defined. On the other
hand, it could be expected that the hidden conformal symmetry should
be still there. In this paper, we try to address this issue. We
introduce a new set of conformal coordinates and compute the
corresponding $SL(2,R)$ quadratic Casimir. We find that for many
extremal black hole, the scalar radial equation could be written as
the $SL(2,R)$ Casimir, showing that   the hidden
conformal symmetries for extremal black holes do exist. We read the dual
temperature from the conformal coordinate directly and find
agreement with the known one in the extremal limit. We furthermore
discuss the scattering amplitudes off the extremal black hole and
find good agreement with the CFT prediction.

In the next section, we introduce the new conformal coordinates and
discuss the hidden conformal symmetry. In Sec. 3, we study the
extremal Kerr(-Newman) black hole. In Sec. 4, we turn to other
extremal black holes, including warped AdS$_3$, null warped, the near-horizon geometry
of extreme Kerr black hole (NHEK)
and 4D RN black hole. We end with some conclusions in Sec. 5.

\section{Hidden conformal symmetry}

For extreme black holes,  the conformal coordinates are very
different from the ones of nonextremal black holes widely used in
the literature. They could be defined as follows \bea
\o^+&=&\frac{1}{2}\left(\a_1 t+\b_1 \phi-\frac{\g_1}{r-r_+}\right),\nn\\
\o^-&=& \frac{1}{2}\left(e^{2\pi T_L\phi+2n_L t}-\frac{2}{\g_1}\right),\label{confcor}\\
y&=&\sqrt{\frac{\g_1}{2(r-r_+)}}e^{\pi T_L\phi+n_Lt}.\nn \eea With
them, we can locally define the vector fields \bea
H_1&=&i\p_+ \nn\\
H_0&=&i\left(\o^+\p_++\frac{1}{2}y\p_y\right) \nn\\
H_{-1}&=&i(\o^{+2}\p_++\o^+y\p_y-y^2\p_-)
\eea
and
\bea
\tilde H_1&=&i\p_- \nn\\
\tilde H_0&=&i\left(\o^-\p_-+\frac{1}{2}y\p_y\right) \nn\\
\tilde H_{-1}&=&i(\o^{-2}\p_-+\o^-y\p_y-y^2\p_+)
\eea
These vector fields obey the $SL(2,R)$ Lie algebra
\be
[H_0, H_{\pm 1}]=\mp iH_{\pm 1},\hs{5ex} [H_{-1},H_1]=-2iH_0,
\ee
and similarly for $(\tilde H_0, \tilde H_{\pm 1})$. The quadratic Casimir is
\bea
\cH^2=\tilde{\cH}^2&=&-H_0^2+\frac{1}{2}(H_1 H_{-1}+H_{-1}H_1) \nn\\
 &=&\frac{1}{4}(y^2\p^2_y-y\p_y)+y^2 \p_+\p_-.
 \eea

 In terms of $(t,r,\phi)$ coordinates, the vector fields are of the form:
 \bea
 H_1&=&i\frac{2}{A}(2\pi T_L\p_t-2n_L\p_\phi)\nn\\
 H_0&=&i\left(-(r-r_+)\p_r+\frac{1}{A}(\a_1t+\b_1\phi)(2\pi
 T_L\p_t-2n_L\p_\phi)\right)\nn\\
 H_{-1}&=&i\left\{-(\a_1t+\b_1\phi)(r-r_+)\p_r+\frac{\g_1}{A(r-r_+)}(\b_1\p_t-\a_1\p_\phi)\right.\nn\\
&&\left. +\left((\a_1t+\b_1\phi)^2
 +\frac{\g_1^2}{(r-r_+)^2}\right)\frac{1}{2A}(2\pi
 T_L\p_t-2n_L\p_\phi)\right\}\nn\\
\tilde H_1&=&2ie^{-2\pi
T_L\phi-2n_Lt}\left((r-r_+)\p_r-\frac{1}{A}(\b_1\p_t-\a_1\p_\phi)-\frac{\g_1}{A(r-r_+)}(2\pi
 T_L\p_t-2n_L\p_\phi)\right)\nn\\
 \tilde H_0&=&i\left(-\frac{2}{\g_1}e^{-2\pi
T_L\phi-2n_Lt}(r-r_+)\p_r-(1-\frac{2}{\g_1}e^{-2\pi
T_L\phi-2n_Lt})\frac{1}{A}(\b_1\p_t-\a_1\p_\phi)\right.\nn\\
 &&\left.
+\frac{2e^{-2\pi T_L\phi-2n_Lt}}{A(r-r_+)}(2\pi
 T_L\p_t-2n_L\p_\phi)\right)\nn\\
\tilde H_{-1}&=&i\left\{-\frac{1}{2}\left(e^{2\pi
T_L\phi+2n_Lt}-\frac{4}{\g^2_1}e^{-2\pi
T_L\phi-2n_Lt}\right)(r-r_+)\p_r \right.\nn\\
& & -\left(e^{2\pi
T_L\phi+2n_Lt}-\frac{4}{\g_1}+\frac{4}{\g^2_1}e^{-2\pi
T_L\phi-2n_Lt}\right)\frac{1}{2A}(\b_1\p_t-\a_1\p_\phi) \nn\\
&&\left. -\left(e^{2\pi T_L\phi+2n_Lt}+\frac{4}{\g^2_1}e^{-2\pi
T_L\phi-2n_Lt}\right)\frac{\g_1}{2A(r-r_+)}(2\pi
 T_L\p_t-2n_L\p_\phi)\right\}\nn
 \eea
and the quadratic Casimir becomes
 \bea
 \cH^2 &=& \p_r(\D \p_r)-\left(\frac{\g_1(2\pi T_L\p_t-2n_L\p_\phi)}{A(r-r_+)}\right)^2
  -\frac{2\g_1(2\pi
  T_L\p_t-2n_L\p_\phi)}{(r-r_+)A^2}(\b_1\p_t-\a_1\p_\phi),
  \label{Casimir}
  \eea
  where $A=2\pi T_L\a_1-2n_L\b_1$ and $\D=(r-r_+)^2$.

Actually, there exist some degrees of freedom to define the
conformal coordinates (\ref{confcor}) without affect the  form of
the Casimir. For example, the $\g_1$ term in $\o^-$ is redundant and
could be abandoned and there could be a free overall scale factor in
$\o^-$. However, such degrees of freedom do not change the
underlying physics.

Similar to the hidden conformal symmetry of the nonextremal Kerr-Newman
black hole, the vector fields are not globally defined. The periodic
identification along $\phi$ breaks the conformal symmetry. As argued
in \cite{Castro:2010fd}, from the relation between $\o^-$ and
$t^-=2\pi T_L+2n_Lt$, one may take $t^-$ as the Rindler coordinate
such that the observer in it will observe a thermal bath of Unruh
radiation with temperature $T_L$. Note that for extremal black
holes, only one sector in the dual CFT is excited.

It turns out that the radial equations of the extremal black holes
always take the form \be \cH^2 \Phi(r)=K\Phi(r), \ee where $K$ is a
$r$-independent parameter contributing  the conformal weight. With
the ansatz $\Phi(r)=e^{-i\o t+im\phi}R(r)$, the radial equation can
be written as \be
\partial_r\Delta\partial_r R(r)+\frac{\mu^2}{(r-r_+)^2}R(r)+\frac{\nu}{r-r_+}R(r)=KR(r), \label{radialgeneral}
\ee
where
\bea
\mu&=&\frac{\gamma_1(2\pi T_L\omega+2n_Lm)}{A}\nn\\
\nu&=&\frac{2\gamma_1(2\pi T_L\omega+2n_L m)(\beta_1\omega+\alpha_1m)}{A^2}.
\eea

Introducing  $z=\frac{-2i\mu}{r-r_+}$, we get the Whittaker equation
\be
R^{''}(z)+(-\frac{1}{4}+\frac{k}{z}+\frac{\frac{1}{4}-m_s^2}{z^2})R(z)=0,
\ee where \be k=\frac{i(\beta_1\omega+\alpha_1m)}{A},
\hs{3ex}m_s^2=\frac{1}{4}+K. \ee This equation has the solution \be
R(z)=C_1R_+(z)+C_2R_-(z), \ee where
$R_\pm(z)=e^{-\frac{z}{2}}z^{\frac{1}{2}\pm m_s}F(\frac{1}{2}\pm
m_s-k,1\pm2m_s,z)$ are two linearly independent solution. Near the
horizon $r\to r_+$ so $z\to\infty$, the Kummer function could be
expanded asymptotically \be F(\alpha,\gamma,z)\sim
\frac{\Gamma(\gamma)}{\Gamma(\gamma-\alpha)}e^{-i\alpha\pi}z^{-\alpha}+\frac{\Gamma(\gamma)}{\Gamma(\alpha)}e^zz^{\alpha-\gamma}.
\ee As we need to impose purely ingoing boundary condition at the
horizon, we have to require \be
C_1=-\frac{\Gamma(1-2m_s)}{\Gamma(\frac{1}{2}-m_s-k)}C,
\hs{3ex}C_2=\frac{\Gamma(1+2m_s)}{\Gamma(\frac{1}{2}+m_s-k)}C \ee to
cancel the outgoing modes, where $C$ is a constant.

When $r$ goes asymptotically to infinity, $z\to 0$,
$F(\alpha,\gamma,z)\to1$, the solution has asymptotical behavior \be
R\sim C_1 r^{-h}+C_2 r^{1-h}, \ee where $h$ is the conformal weight
of the scalar \be
h=\frac{1}{2}+m_s=\frac{1}{2}+\sqrt{\frac{1}{4}+K}. \ee The retarded
Green's function could be read directly\cite{Son05} \be
G_R\sim\frac{C_1}{C_2}\propto\frac{\Gamma(1-2h)\Gamma(h-k)}{\Gamma(2h-1)\Gamma(1-h-k)}.
\ee

\section{Extremal Kerr-Newman black hole}

For the Kerr-Newman black hole with mass $M$, angular momentum
$J=aM$ and electric charge $Q$, its metric takes the following form
\be \label{KerrNewman}
ds^2=-\frac{\D}{\r^2}(d t-a\sin^2\th d\phi)^2+
\frac{\r^2}{\D}dr^2+\rho^2 d\th^2+ \frac{1}{\r^2}\sin^2\th\left(ad
t-(r^2+a^2)d\phi\right)^2, \ee where \bea
\Delta&=&(r^2+a^2)-2Mr+Q^2, \nn\\
\r^2&=&r^2+a^2\cos^2\th.
\eea
The gauge field is \be A=-\frac{Qr}{\r^2}(d t-a\sin^2\th d\phi).
\ee There are two horizons located at \be r_\pm=M\pm
\sqrt{M^2-a^2-Q^2}. \ee And the Hawking temperature, entropy,
angular velocity of the horizon and the electric potential are
respectively
 \bea
 T_H&=&\frac{r_+-r_-}{4\pi(r_+^2+a^2)},\nn\\
 S&=&\pi(r_+^2+a^2),\nn\\
 \O_H&=&\frac{a}{r_+^2+a^2},\nn\\
 \Phi&=&\frac{Qr_+}{r_+^2+a^2}.
 \eea

Let us consider the complex scalar field with mass $\mu$ and charge $e$ scattering with
the Kerr-Newman black hole. The Klein-Gordon equation is
\be
(\nabla_\mu+ieA_\mu)(\nabla^\mu+ieA^\mu)\Phi-\mu^2\Phi=0.
\ee
With the ansatz
\be\label{ansatz}
 \Phi=e^{-i\o t+im\phi}\cR(r)\cS(\th),
 \ee
where $\o$ and $m$ are the quantum numbers, the wave equation could
be decomposed into the angular part and the radial part. The angular
part is of the form \be
 \frac{1}{\sin\th}\frac{d}{d\th}\left(\sin\th\frac{d}{d\th}\cS\right)+\left(
 \L_{lm}-a^2(\o^2-\mu^2)\sin^2\th-\frac{m^2}{\sin^2\th}\right)\cS=0.
 \ee
Here $\L_{lm}$ is the separation constant. It is
restricted by the regularity
boundary condition at $\th=0,\pi$ and can be computed
numerically. The radial part of the wave function is of the form
\be
\p_r(\D \p_r\cR)+V_R \cR=0
\ee
with
\bea
V_R&=&-\L_{lm}+2am\o+\frac{H^2}{\D}-\m^2(r^2+a^2), \\
H&=&\o(r^2+a^2)-eQr-am.
\eea

As we are interested in the low-frequency limit, \be\label{low} \o
M <<1, \ee the $\o^2$ term in the angular equation could be
neglected. Note that the low-frequency limit (\ref{low}) is very
different from the case studied in
\cite{Bredberg:2009pv,{Hartman:2009nz}}, where only the
frequencies near the superradiant bound were studied. To simplify
our discussion, we focus on the massless scalar. Then the angular
equation is just the Laplacian on the 2-sphere with the separation
constants taking values \be \L_{lm}=l(l+1). \ee

In the near region, $r\o <<1$, the radial equation could be simplified even more
\bea
\lefteqn{\p_r\D \p_r\cR(r)+\frac{(ma-\o(2Mr_+-Q^2)+eQr_+)^2}{(r-r_+)(r_+-r_-)}\cR(r)}\nn\\
&&-\frac{(ma-\o(2Mr_--Q^2)+eQr_-)^2}{(r_+-r_-)(r-r_-)}\cR(r)=\left(l(l+1)-e^2Q^2\right)\cR(r)\label{radial}
\eea

Let us consider the extreme Kerr-Newman black hole. In this case,
the Hawking temperature is vanishing but the entropy and the angular
velocity is still nonzero. In the low-frequency limit, we have the
radial equation  as \bea
\lefteqn{\p_r\D \p_r\cR(r)+\frac{2(2M\o-eQ)((2Mr_+-Q^2)\o-am-eQr_+)}{(r-r_+)}\cR(r)}\nn\\
&&+\frac{((2Mr_+-Q^2)\o-am-eQr_+)^2}{(r-r_+)^2}\cR(r)=\left(l(l+1)-e^2Q^2\right)\cR(r).\label{radial}
\eea
For the neutral scalar with $e=0$, it is easy to see that the left
hand side of this equation could be rewritten as the $SL(2,R)$
Casimir (\ref{Casimir}) with the identification
  \be
  \a_1=0, \hs{3ex} \b_1=\g_1/a, \hs{3ex}2\pi T_L=\frac{2r_+-Q^2/M}{2a},\hs{3ex}
  n_L=-\frac{1}{4M}.
  \ee
  The identifications of $T_L$ and $n_L$ are consist with the existing result\cite{Chen:2010xu,{Wang:2010qv}}.

\subsection{Microscopic description}

 Now the microscopic entropy comes from only the left sector
 \be
 S=\frac{\pi^2}{3}c_LT_L=\pi (r_+^2+a^2),
 \ee
 in agreement with the macroscopic Bekenstein-Hawking entropy.

 We can still determine the conjugate charge from the first law of thermodynamics.
  In the extreme case, the Hawking temperature $T_H=0$, we should consider it carefully.
  We begin with a nonzero $T_H$, and then take limit to set it to $0$. When $T_H\neq0$,
  \be
M^2-a^2-Q^2>0. \ee From the first law of thermodynamics, \be \delta
S=\frac{\delta M-\Omega_H\delta J-\Phi \delta Q}{T_H} \ee we get \be
\delta S=2\pi (2M\delta M-Q\delta Q)+4\pi\frac{(2M^2-Q^2)\delta
M-a\delta J-QM\delta Q}{2\sqrt{M^2-a^2-Q^2}} \ee To analyze the
second term, we introduce \be
 a=M(1-\epsilon)\cos\theta,\hs{3ex} Q=M(1-\epsilon)\sin\theta
\ee where $\epsilon$ and $\theta$ are two parameter to control the
 $T_H\to0$ limit. Note that only $\epsilon$ is really related to the
 limit
$T_H\to0$, so we may change $M$ and $\theta$ simultaneously: \bea
\delta J&=&2M(1-\epsilon)\cos\theta\delta
M-M^2(1-\epsilon)\sin\theta\delta\theta, \nn\\
 \delta
Q&=&(1-\epsilon)\sin\theta\delta M+M(1-\epsilon)\cos\theta
\delta\theta, \nn\eea then \be \frac{(2M^2-Q^2)\delta M-a\delta
J-QM\delta Q}{2\sqrt{M^2-a^2-Q^2}}=M\delta M\sqrt{1-(1-\epsilon)^2}.
\ee In the limit $T_H\to0$ the second term turns to zero.

Next, we consider the first term. We find that if we  identify \be
\delta Q=e,\hs{3ex}\delta M=\omega,\hs{3ex} \delta
E_L=\omega_L-q_L\mu_L,\ee with \be
\omega_L=\frac{(2M^2-Q^2)M}{J}\omega,\hs{3ex}\mu_L=\frac{QM^2}{J}-\frac{Q^3}{2J},\hs{3ex}q_L=e,
\label{identi}\ee then \be \delta S=\frac{\delta
E_L}{T_L}=\frac{\omega_L-q_L\mu_L}{T_L}. \ee  Note that the
identification (\ref{identi}) is the same as the one found in the
study of the nonextremal Kerr-Newmann black hole. The above discussion
also suggests that only the left mover in the dual CFT is relevant
to the extremal black hole, while the right mover is  kept completely 
silent.

The retarded charged scalar Green's function in the extremal Kerr-Newman
black hole could be rewritten as
 \bea
 G_R &\sim
 &\frac{\Gamma(1-2h)\Gamma(h-i(2M\o-eQ))}{\Gamma(2h-1)\Gamma(1-h-i(2M\o-eQ))}\nn\\
  &=&\frac{\Gamma(1-2h)}{\Gamma(2h-1)}\frac{\Gamma\left(h-i\frac{\o_L-q_L\mu_L}{2\pi T_L}\right)}
  {\Gamma\left(1-h-i\frac{\o_L-q_L\mu_L}{2\pi T_L}\right)}.
  \label{KNcorrelator}
\eea Now the conformal weight
$h=\frac{1}{2}+\sqrt{\frac{1}{4}+l(l+1)-e^2Q^2}$.

In a two-dimensional conformal field theory, the two-point functions of the primary operators are determined by the
conformal invariance\cite{Cardy:1984bb}. The retarded correlator  $G_R (\o_L, \o_R)$ is analytic on the
upper half complex $\o_{L,R}$ plane and its value along the
positive imaginary $\o_{L,R}$ axis gives the Euclidean correlator:
\be \label{GER} G_E(\o_{L,E}, \o_{R,E}) = G_R(i\o_{L,E},
i\o_{R,E}), \quad \o_{L,E} , \o_{R,E} >0. \ee At finite temperature, $\o_{L,E}$ and $\o_{R,E}$ take discrete
values of the Matsubara frequencies \be \o_{L,E} =  2 \pi m_L T_L,
\quad \o_{R,E} =  2 \pi m_R T_R, \ee where $m_L, m_R$ are integers
for bosonic modes and are half integers for fermionic modes. For an operator of
dimensions $(h_L,h_R)$, charges $(q_L,q_R)$ at temperatures
$(T_L,T_R)$ and chemical potentials $(\m_L, \m_R)$, the momentum space Euclidean correlator is given
by\cite{Maldacena:1997ih}
 \bea \label{GE}
G_E(\o_{L,E}, \o_{R,E}) &\sim& T_L^{2 h_L-1}  T_R^{2 h_R-1} e^{i
\frac{\bo_{L,E}}{2T_L}} e^{i \frac{\bo_{R,E}}{2T_R}}\nn\\
&&\cdot\G(h_L + \frac{\bo_{L,E}}{2 \pi T_L})\G(h_L -
\frac{\bo_{L,E}}{2 \pi T_L}) \G(h_R + \frac{\bo_{R,E}}{2 \pi
T_R})\G(h_R - \frac{\bo_{R,E}}{2 \pi T_R}), \eea where \be
\bo_{L,E}= \o_{L,E} - i q_L \m_L, \quad \bo_{R,E}= \o_{R,E} - i
q_R \m_R. \ee

In our case, only the left part of the CFT is necessary, then
\be\label{GEleft} G_E(\o_{L,E})\sim T_L^{2 h_L-1}e^{i
\frac{\bo_{L,E}}{2T_L}}\G(h_L + \frac{\bo_{L,E}}{2 \pi T_L})\G(h_L -
\frac{\bo_{L,E}}{2 \pi T_L}). \ee
 The real-time correlator (\ref{KNcorrelator}) is obviously in
 agreement with the above relation.

\subsection{Kerr case}

When $Q=0$, the Kerr-Newman black hole reduces to the Kerr black
hole. For the extreme Kerr, we have \be a=M, \hs{3ex}
T_L=\frac{1}{2\pi}, \hs{3ex} \o_L=2M\o. \ee

In this case, we can study the scattering amplitudes for general
spin $s$\cite{Hartman:2009nz,{Chen:2010xu}}. The radial equation is
\be
\Delta^{-s}\partial_r\Delta^{s+1}\partial_rR(r)+[\frac{H^2-2is(r-r_+)}{\Delta}+4is\omega
r+2ma\omega+s(s+1)-K]R(r)=0 \ee where $H=\omega(r^2+a^2)-ma$ and $K$
is the separation constant. When we consider the low-frequency
limit, we get the following equation \be
\Delta^{-s}\partial_r\Delta^{s+1}\partial_rR(r)+(s(s+1)-K)R(r)+\frac{d}{r-r_+}R(r)+\frac{e^2}{(r-r_+)^2}R(r)=0
\ee where \bea
d&=&(4M\omega-2is)(2M\omega r_+-ma),\nn\\
e&=&2M\omega r_+-ma. \nn \eea Setting $R(r)=(r-r_+)^{-s}\psi(r)$, we
get the following equation \be
\partial_r\Delta\partial_r\psi+\frac{d}{r-r_+}\psi+\frac{e^2}{(r-r_+)^2}\psi=K\psi,
\ee which is of the same as the Eq. (\ref{radialgeneral}). The
retarded Green's function could be obtained in the way suggested in
\cite{ChenChu}. The result is \be G\sim
\frac{\G(1-2h)\G(h_L-i\frac{\o_L}{2\pi
T_L})}{\G(2h-1)\G(1-h_L-i\frac{\o_L}{2\pi T_L})} \ee where
$h_L=h-s$. This is in agreement with the CFT prediction.

\section{Other extreme black holes}

There has been much discussion on the hidden conformal symmetry
acting on the solution space of the scalar field in other black
holes. We will show that for the extremal cases, there exist hidden
conformal symmetry as well. In all the cases we study, the
conformal coordinates proposed in  Sec. 2 work well.

\subsection{Extreme warped AdS$_3$ black hole}

The spacelike stretched $AdS_3$ spacetime is the vacuum solution of
three-dimensional topological massive
gravity\cite{Deser:1981wh,Deser:1982vy}. The spacelike stretched
$AdS_3$ black hole could be constructed from discrete identification
along a Killing vector of the global warped spacetime, similar to
the BTZ black hole\cite{BTZ}.

The metric of the spacelike stretched warped $AdS_3$ black hole
takes the following form in terms of Schwarzschild coordinates:
  \bea
    ds^2=l^2(\textrm{d}t^2+2M(r)\textrm{d}t\textrm{d}\theta+N(r)\textrm{d}\theta^2+D(r)\textrm{d}r^2),
 \eea
where \bea
 M(r)&=& v r-\frac{1}{2}\sqrt{r_+r_-(v^2+3)},\nn\\
 N(r)&=&\frac{r}{4}\left(3(v^2-1)r+(v^2+3)(r_++r_-)-4v\sqrt{r_+r_-(v^2+3)}\right),\nn\\
 D(r)&=&\frac{1}{(v^2+3)(r-r_+)(r-r_-)},\nn
 \eea
and $-l^{-2}$ is the negative cosmological constant and the
parameter
 $v=\mu l/3$ with $\mu$ being the mass of the graviton.  There are two horizons located at
 $r=r_+$ and $r=r_-$. It is only well behaved for $\nu>1$. The hidden conformal symmetry
 in this black hole has been discussed in \cite{Fareghbal:2010yd}.

 From the warped AdS/CFT correspondence\cite{Andy08}, this black hole could be described by a two-dimensional conformal field theory with temperatures
 \bea\label{tempwarped}
  T_L&=&\frac{(v^2+3)}{8\pi
  l}\left(r_++r_--\frac{\sqrt{(v^2+3)r_+r_-}}{v}\right), \\
  T_R&=&\frac{(v^2+3)(r_+-r_-)}{8\pi
  l},
 \eea
 and central charges
 \be
  c_L=\frac{l}{G}\frac{4v}{v^2+3}, \hspace{5ex}
  c_R=\frac{l}{G}\frac{5v^2+3}{v(v^2+3)}.
  \ee

 The scalar field of mass $\mu$ in this background obeys the equation
of motion: \be
  (\nabla_\nu\nabla^\nu-\mu^2) \Phi=0.
  \ee
  Since the background has the translational isometry along $t$
  and $\th$, we may make the following ansatz
  \be
  \Phi=e^{-i\o t+ik\th}\phi.
  \ee
  For the extremal warped black hole, the scalar equation turns to be
  \bea
\lefteqn{\p_r\D \p_r\phi(r)+\frac{4v\o((2v-\sqrt{v^2+3})r_+\o+2k)}{(r-r_+)(v^2+3)^2}\phi(r)}\nn\\
&&+\frac{((2v-\sqrt{v^2+3})r_+\o+2k)^2}{(r-r_+)^2(v^2+3)^2}\phi(r)=\left(\frac{\mu^2l^2}{v^2+3}-\frac{3(v^2-1)}{(v^2+3)^2}\o^2\right)\phi(r)\label{radial}
\eea Note that the right hand side is closely related to the
conformal weight of the scalar field, taken into account of the
quantum number identification.  For the identification of quantum
numbers, please see \cite{{ChenXu2},Chen:2009cg} for details. One
does not need to take the low-frequency limit as done in
\cite{Fareghbal:2010yd}, which makes the warped AdS/CFT
correspondence less clear. Moreover, the hidden conformal symmetry
exist in the whole spacetime, rather than just the ``Near" region in
the  Kerr case. The left hand side of the equation could be written
as the $SL(2,R)$ quadratic Casimir (\ref{Casimir}) with the
identifications \bea
&& \a_1=0, \hs{3ex}\b_1=\frac{v^2+3}{2} \g_1 \nn\\
&& 2\pi T_L=\frac{v^2+3}{4vl}(2v-\sqrt{v^2+3})r_+, \hs{3ex} n_L=\frac{v^2+3}{4vl}.
\eea
The left temperature is exactly the same as the one in (\ref{tempwarped}). Similar to
the Kerr/CFT case, for the extreme black holes, only the left temperature is nonvanishing.
The scattering amplitude could be discussed straightforwardly, taking care of the subtle identification
of quantum numbers found in \cite{ChenXu09,{ChenXu2}}. The result is in perfect with the CFT
prediction.

\subsection{Null warped black holes}

The null warped $AdS_3$ spacetime is another vacuum solution of
three-dimensional topological massive gravity. It is only well
defined at $v=1$. The null warped black hole could be taken as the
quotient of the null warped $AdS_3$. The  metric of the null warped
black hole is of the form
 \be
   \frac{ds^2}{l^2}=-2r\textrm{d}\theta\textrm{d} t+(r^2+r+\alpha^2)\textrm{d}
   \theta^2+\frac{\textrm{d} r^2}{4r^2},
 \ee
 where $1/2>\alpha>0$ in order to avoid the naked causal
 singularity. The horizon is located at $r=0$. Different from its spacelike stretched
 cousin, this black hole is extreme by construction. From the null warped AdS/CFT correspondence\cite{Andy08,{ChenXu2}}, it was suggested to be described by
a dual two-dimensional CFT with only nonvanishing right-moving temperature
 \be
 T_R=\frac{\a}{\pi l}
 \ee
and the central charge\cite{Anninos:2010pm}
 \be C_R=2l/G. \ee In
this case, there is only right central charge in the dual CFT,
suggesting that the dual CFT is chiral.

The equation of
motion for the scalar field of mass $\mu$ is
 \be
   \nabla^2\Phi-\mu^2\Phi=0.
\ee
  Taken the ansatz $\Phi=e^{-i\omega t+ik\theta}R(r)$, the
  equation
 becomes
 \be
   \frac{\textrm{d} }{\textrm{d} r}(r^2\frac{\textrm{d}}{\textrm{d} r}R)+\left(\frac{\o^2-2\omega
   k}{4r}+\frac{\omega^2\alpha^2}{4r^2}\right)R=\frac{\mu^2l^2-\o^2}{4}R.
 \ee
 Similar to the spacelike warped black hole, the right hand side of the equation is related to
 the conformal weight of the scalar field. The left hand side of the equation could be written as the quadratic Casimir (\ref{Casimir}) with the following identification
 \be
 \a_1=-2\b_1, \hs{3ex}\g_1=-\frac{\a}{2}\a_1, \hs{3ex}
 n_L=0, \hs{3ex} T_L=\frac{\a}{\pi l}.
 \ee
The scattering amplitudes of null warped black hole have been discussed in detail in \cite{Chen:2009cg}.

 \subsection{NHEK}

The near-horizon geometry of extreme Kerr black hole (NHEK) was first discovered in \cite{Bardeen:1999px}. It
played a key role in setting up the Kerr/CFT correspondence.
 The NHEK geometry in Poincare-type coordinates is
 \be\label{NHEK}
 ds^2=\G\left(-r^2dt^2 +
 \frac{dr^2}{r^2}+d\th^2+\L^2(d\phi+rdt)^2\right).
 \ee
 where \be
 \G(\th)=\frac{1+\cos^2\th}{2},
 \hs{3ex}\L(\th)=\frac{2\sin\th}{1+\cos^2\th}.
 \ee
 It was shown in \cite{ChenChu} that the NHEK geometry could be taken as the extreme black hole,
 with the horizon located at $r=0$.

 With the ansatz \be
 \Phi =e^{-i\o t+im\phi}\cR(r)\cS(\th),
 \ee
  the radial  function obeys
 \be
 \frac{d}{dr}\left(r^2\frac{d}{dr}\right)\cR(r)+\left(\frac{\o^2}{r^2}+\frac{2\o m}{r}\right)\cR(r)=\left(
 \L_{lm}-2m^2\right)\cR(r).
 \ee
The coefficient in the right hand side is related to the conformal
weight. The left hand side can be recast into the quadratic Casimir
(\ref{Casimir}) with the following identification \be \b_1=0,
\hs{3ex} \g_1=\a_1 \hs{3ex} n_L=0, \hs{3ex} T_L =\frac{1}{2\pi}. \ee
The temperature is exactly the same as the one found in
\cite{AndyWei}. The scattering amplitudes for various kinds of
perturbations in NHEK have been discussed in \cite{ChenChu}.

A similar discussion  could  easily be applied to the extremal self-dual
warped black hole\cite{Chen:2010qm}.

\subsection{Extreme RN black hole}

The four-dimensional electrically charged RN black hole is described by the metric
\be ds^2=-f(r)dt^2+\frac{dr^2}{f(r)}+r^2d\O^2, \ee with
 \be
f(r)=1-\frac{2M}{r}+\frac{Q^2}{r^2} \ee and the gauge potential \be
A=-\frac{2Q}{r}dt. \ee The Hawking temperature and entropy are
respectively \be T_H=\frac{r_+-r_-}{4\pi r^2_+}, \hs{3ex}S=\pi
r^2_+, \ee where $r_\pm=M\pm \sqrt{M^2-Q^2}$.  In order to study the
CFT dual of the four-dimensional RN black hole, one has to embed it into five
dimension. From the RN/CFT correspondence\cite{Hartman:2008pb}, it
is described by a dual CFT with central charges \be c_L=c_R=6Q^3,
\ee and temperatures \be T_L=\frac{(r_++r_-)M}{2\pi
Q^3}-\frac{1}{2\pi Q}, \hs{3ex} T_R=\frac{(r_+-r_-)M}{2\pi Q^3}. \ee

For the generic nonextremal RN black hole, its hidden conformal
symmetry has been discussed in \cite{Chen:2010as}. For the extremal
RN black hole,  the scalar radial equation in the low-frequency and
low momentum limit takes the form \be
\p_r(\D\p_rR(r)+\frac{2M^3(\o-m))(2\o-m)}{r-r_+}R(r)+\frac{M^4(\o-m)^2}{(r-r_+)^2}R(r)=l(l+1)R,
\ee where $\o$ and $m$ are the quantum numbers corresponding to the
translation along the time and the fifth dimension. In this case, it
is easy to see the equation could be rewritten as the $SL(2,R)$
Casimir (\ref{Casimir}) with the identification \bea
\a_1=-aM, &&\b_1=2aM, \hs{3ex}\g_1=M^3a \nn\\
n_L=-\frac{1}{2M}, & & ~~2\pi T_L=\frac{1}{M} \eea where $a$ is just
a free parameter. Once again, the left temperature is in consistency
with the one found in generic case. The scattering amplitude could
be computed in a similar way and is in agreement with the CFT
prediction.

\section{Conclusion}

In this paper, we studied the hidden conformal symmetry of extremal
Kerr black holes. We introduced a new set of conformal coordinates
(\ref{confcor}), which include five parameters
$(\a_1,\b_1,\g_1,T_L,n_L)$. It is nice to see that the induced
$SL(2,R)$ quadratic Casimir is capable of rewriting the scalar
Laplacian for various kinds of extremal black holes. The parameter
$T_L$ has clear physical meaning, corresponding to the temperature
of the dual CFT. The other parameters take different values for
various kinds of black hole and there is always one more
undetermined degree of freedom among $(\a_1,\b_1,\g_1)$. Note also
the fact that all the black holes we discussed in this paper have
holographic two-dimensional CFT descriptions is in accordance with the existence
of hidden conformal symmetry in these black holes. We believe that
this should be true for other extremal black holes with holographic
pictures\cite{Astefanesei:2009sh}.

 The microscopic descriptions of these black holes
fit very nicely into their established holographic pictures,
including Kerr/CFT, warped AdS/CFT, and RN/CFT et.al.. The temperatures
we identified are in agreement with the ones found in the literature
taking extremal limit. The Bekenstein-Hawking entropy and the
scattering amplitudes are all in agreement with the CFT prediction.
For the extremal black holes, their dual CFT is ``chiral" in the
sense that only one sector has nonvanishing temperature and the
other sector is completely cooled down without excitation. This fact
is  already reflected in the conformal coordinates, in which only the
left-moving temperature appears and the right-moving temperature is
set to zero by definition.

\section*{Acknowledgments}

 This research was  supported in part by NSFC Grant
Nos.10775002,10975005, NKBRPC (No. 2006CB805905), and the Project of
Knowledge Innovation Program (PKIP) of the Chinese Academy of Sciences under 
Grant No. KJCX2.YW.W10. BC would like to thank KITPC for
hospitality, where part of the work was done.

\end{document}